\documentclass[a4paper]{jpconf}
\usepackage{graphicx}

\usepackage{color} % Für Korrekturzwecke

\begin{document}

\title{The GEO\,600 squeezed light source}

\author{Henning Vahlbruch, Alexander Khalaidovski, Nico Lastzka, Christian Gr\"af, Karsten Danzmann, and Roman Schnabel}
\address{Institut f\"ur Gravitationsphysik of Leibniz Universit\"at Hannover and Max-Planck-Institut f\"ur Gravitationsphysik (Albert-Einstein-Institut), Callinstr. 38, 30167 Hannover, Germany}
\ead{roman.schnabel@aei.mpg.de}

\begin{abstract}
The next upgrade of the GEO\,600 gravitational wave detector is scheduled for 2010 and will, in particular, involve the implementation of  squeezed light. The required non-classical light source is assembled on a 1.5\,m$^2$ breadboard and includes a full coherent control system and a diagnostic balanced homodyne detector. Here, we present the first experimental characterization of this setup as well as a detailed description of its optical layout. A squeezed quantum noise of up to 9\,dB below the shot-noise level was observed in the detection band between 10\,Hz and 10\,kHz. We also present an analysis of the optical loss in our experiment and provide an estimation of the possible non-classical sensitivity improvement of the future squeezed light enhanced GEO\,600 detector.
\end{abstract}

\section{Introduction}

Photon shot-noise is a limiting noise source in laser interferometric gravitational wave (GW) detectors. The signal to shot-noise ratio can be improved by increasing the laser power. For this reason the planned Advanced LIGO detectors are designed to store about a megawatt of optical power inside the interferometer arms \cite{SMITH09}. At such high laser powers thermally induced optical waveform distortion due to light absorption and the excitation of parasitic instabilities might become an issue \cite{Dambr03, Shaug04}. Alternatively, the signal to shot-noise ratio can also be improved by `squeezing' the shot-noise as proposed by Caves in 1981 \cite{Cav81}. In this case, the laser power inside the interferometer is not increased. In order to squeeze the shot-noise of a Michelson interferometer that is operated close to a dark fringe, squeezed (vacuum) states of light have to be injected into the signal output port. A squeezed state is a quantum state whose uncertainty in one of the field quadratures is reduced compared to the vacuum state, while the noise in the conjugate quadrature is increased. Later it was realized that squeezed states of light can also be used to reduce the overall quantum noise in interferometers including radiation pressure noise, thereby beating the standard-quantum-limit (SQL) \cite{Unruh82,JRe90}. Theoretical analysis of Gea-Banacloche and Leuchs \cite{GLe87} and Harms \textit{et al.}\,\cite{HCCFVDS03,SHSD04} suggested that squeezing is broadly compatible with interferometer recycling techniques \cite{DHKHFMW83pr,Mee88} thereby further promoting the application of squeezed states in GW-detectors.

The first observation of squeezed states was done by Slusher \textit{et al.}\,\cite{SHYMV85} in 1985. Since then different techniques for the generation of squeezed light have evolved. One of the most successful approaches for squeezed light generation is optical parametric amplification (OPA), also called parametric down-conversion, based on second-order nonlinear crystals. Common materials like MgO:LiNbO$_3$ or periodicly poled potassium titanyl phosphate (PPKTP) can be used to produce squeezing at the carrier wavelength of today's  GW-detectors operating at 1064 nm. Ground based detectors require a broadband squeezed field in the detection band from about 10 Hz up to 10 kHz. Squeezed states were combined in table top experiments with interferometer recycling techniques \cite{KSMBL02,VCHFDS05}, were demonstrated at audio frequencies \cite{MGBWGCL04, MMGLGGMC05, VCHFDS06, VCDS07}, and were tested on a suspended GW prototype interferometer \cite{GODA08}. A coherent control scheme for the generation and stable control of squeezed vacuum states was also developed for the application in GW-detectors \cite{VCHFDS06, CVDS07}. The progress achieved, now provides all the techniques for a first squeezed light upgrade of a large-scale signal-recycled gravitational wave detector at shot-noise limited frequencies as envisaged in \cite{SHSD04}.

In this paper we present a detailed description of the recently assembled GEO\,600 squeezed light source. The optical layout and also elements of the control scheme  are presented. Our first measurements demonstrate up to 9\,dB squeezing over the complete detection bandwidth of ground-based GW-detectors.

\begin{figure}[t]
\centerline{\includegraphics[width=1\textwidth,keepaspectratio,angle=0]{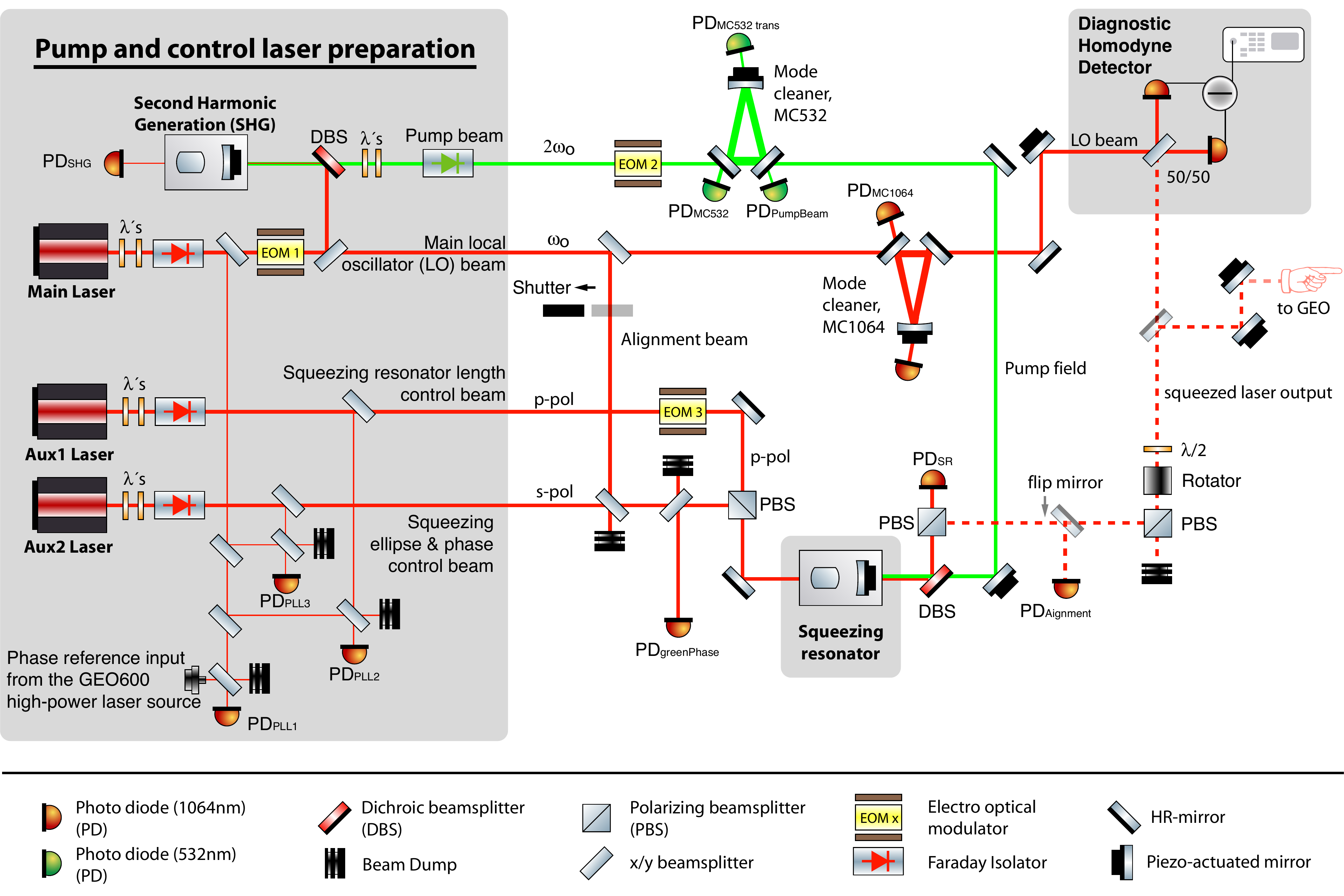}}
\caption{Schematic of the complete optical setup of the GEO\,600 squeezed light source including control lasers and diagnostic tools. Squeezed states of light at audio-band Fourier frequencies around the carrier frequency $\omega_0$ are produced inside a nonlinear squeezing resonator via parametric down-conversion. The squeezing resonator contains a birefringent nonlinear crystal that is pumped with a green laser beam at 532\,nm. Two auxiliary laser beams are also focussed into the crystal and serve as control beams for piezo-electric length stabilization of the cavity and the green pump phase. A balanced homodyne detector was used to characterize the squeezed light output.}
\label{GeneralSetup}
\end{figure}

\section{Experimental}

\begin{figure}[t]
\centerline{\includegraphics[width=1\textwidth,keepaspectratio,angle=0]{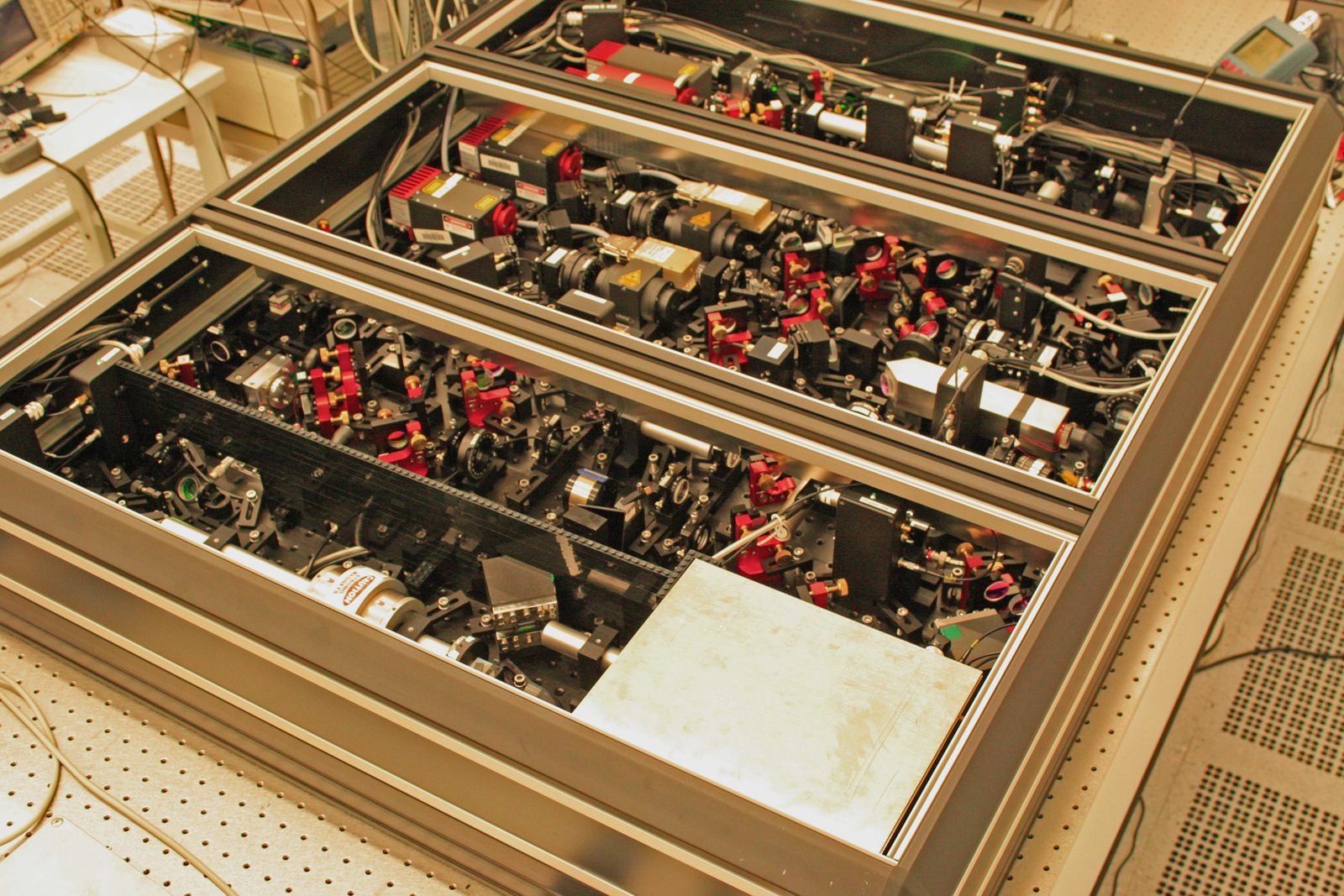}}
\caption{Photograph of the GEO\,600 squeezed light source. The breadboard dimensions are 135\,cm x 113\,cm. The three Nd:YAG Lasers are located on the upper left, at the bottom left the squeezing resonator and on the bottom right the homodyne detector with its covering box is shown. The total weight of the complete system is approximately 130\,kg.}
\label{GEOSqueezerBreadboard}
\end{figure}

The schematic of the optical setup for the GEO\,600 squeezed light source is shown in Figure \ref{GeneralSetup}. For clarity  optical components like lenses, wave plates and steering mirrors etc. are omitted. The optical setup is partly based on earlier experiments presented in \cite{VCHFDS06}-\nocite{VCDS07, CVDS07}\cite{VMCHFLGDS08}. The device has been set up in a class 100 cleanroom in order to avoid dust particles and light scattering. Most optical components like steering mirrors and lenses have super-polished surfaces. The custom-made breadboard of the GEO\,600 squeezed light source has the dimensions of 135\,cm x 113\,cm. A thickness of about 5\,cm (2 inches) in combination with a steel bottom plate and an aluminum inner structure and top plate was chosen to provide a high mechanical stability at a reasonable weight (approximately 70\,kg). The dimensions of the breadboard are adapted to the available space on the GEO\,600 detection bench where the squeezed light source will be operated. The now fully assembled squeezed light source, as shown in Fig.~\ref{GEOSqueezerBreadboard}, has a weight of about 130\,kg.

\begin{figure}[t]
\centerline{\includegraphics[width=0.6\textwidth,keepaspectratio,angle=0]{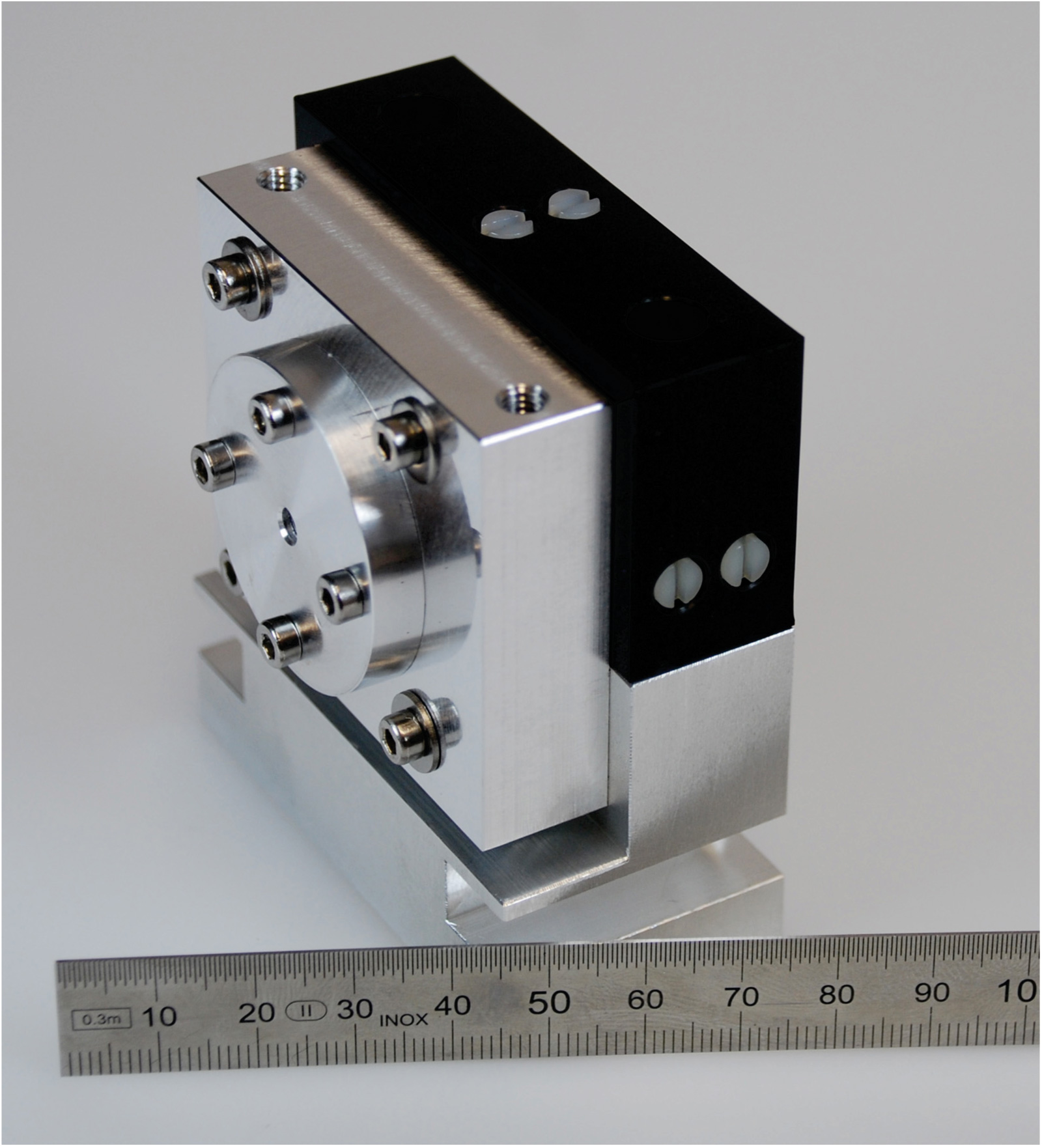}}
\caption{Photograph showing the housing of the nonlinear squeezing resonator. Nonlinear crystal, peltier and piezo-elements as well as the out-coupling mirror are embedded in a compact quasi-monolithic design for a high intrinsic mechanical stability.}
\label{Ofen}
\end{figure}

\subsection{Squeezed-light laser resonator}

The squeezed-light laser resonator (`squeezing resonator' in short) was set up as a standing-wave hemilithic cavity containing a plano-convex PPKTP crystal of about 10\,mm length. The convex crystal surface is high-reflectivity coated for the fundamental and second-harmonic wavelengths at 1064\,nm and 532\,nm, respectively. The planar crystal surface is anti-reflection coated for both wavelengths. The coupling mirror of the standing-wave cavity is a piezo-actuated external mirror with a power reflectivity of R=92\,\% at 1064\,nm and is placed in front of the planar crystal surface at a distance of approximately 20\,mm. The resonator is singly resonant and has a Finesse of about 75 at 1064\,nm. The PPKTP crystal is temperature stabilized to the phase matching temperature of the fundamental, down-converted squeezed (vacuum) field and the second harmonic pump field. Figure \ref{Ofen} shows a photograph of the quasi-monolithic squeezing resonator housing accommodating the nonlinear crystal, peltier and piezo elements as well as the out-coupling mirror.

\subsection{Diagnostic homodyne detector}

The GEO\,600 squeezed light source features an integrated diagnostic balanced homodyne detector (BHD) which could be used to characterize the performance of the squeezed light decoupled from the GEO\,600 interferometer. The required local oscillator beam of about 500\,$\mu$W at 1064\,nm is picked off from the main Nd:YAG laser (2 W, \textit{Mephisto} from Innolight) and is injected into a spatial mode cleaning traveling-wave resonator. This cavity is held on resonance using the PDH-technique at a RF-modulation frequency of 76.5\,MHz with a control bandwidth of 10\,kHz. The transmitted beam interferes with the squeezed beam on a 50/50 beamsplitter with a fringe visibility of 98.6\,\%. Each beamsplitter output field is detected with a single high quantum efficiency photodiode. Both photo currents are subtracted from each other before electronic signal amplification.

\subsection{Preparation of pump and control laser fields}

The GEO\,600 squeezed light source requires altogether four different laser frequencies as illustrated in Figure \ref{GeneralSetup}. A main laser at 1064\,nm provides the optical reference frequency $\omega_0$. Two more laser fields at about 1064\,nm have frequency offsets of more than 10\,MHz with respect to $\omega_0$ and serves as optical control fields. The fourth laser provides the second-harmonic pump field for the parametric down-conversion process at precisely $2\omega_0$. All  four laser fields are mode-matched and injected into the squeezing resonator. We now discuss the preparation and purpose of the four laser fields in more detail.

\textit{Main local oscillator beam at fundamental frequency $\omega_0$} - The main laser source at frequency $\omega_0$ will finally be phase locked to the GEO\,600 laser source. This main laser provides the input field for the second harmonic generator and also serves as a local oscillator beam for homodyne detection of the generated squeezed vacuum field. Additionally, a small fraction of this laser beam is used for the initial alignment of the squeezing resonator. Subsequently this resonator«s length is stabilized using the light transmitted through the resonator, which is detected on the photodiode PD$_{Alignment}$ (see Figure \ref{GeneralSetup}). A RF-demodulation scheme deliveres a squeezing resonator length error signal, which is fed back onto the piezo-actuated cavity out-coupling mirror. 
If an intense alignment beam (about 100\,mW) is injected, a sufficient amount of infrared photons at 1064\,nm gets frequency doubled in order to align the green pump path in the counter direction comfortably. This counter propagating green field (compared to the pump beam) is monitored with either the photodiode PD$_{PumpBeam}$ or PD$_{MC532trans}$. This offers an ideal alignment procedure of this beam path, which could otherwise only be adjusted by measuring  the parametric gain inside the squeezed light source. This latter method can be quite imprecise and time consuming. Nevertheless, the parametric gain can alternatively be monitored using the DC-output of the photodetector PD$_{Alignment}$.
Please note that during the use of this alignment beam only squeezed states at Fourier frequencies in the MHz regime can be generated due to the technical laser noise carried with this beam at lower Fourier frequencies. Therefore, the alignment beam was switched off after the described alignment procedure.

\textit{Second harmonic pump field} - The main laser frequency is frequency doubled by employing a second harmonic generator (SHG). This SHG  produces the necessary pump field for the squeezing resonator at the frequency $2\omega_0$. The design of the SHG is almost identical to the squeezed light source except that  the nonlinear medium was MgO:LiNbO$_3$ instead of PPKTP.  About 35\,mW of pump power are required to achieve an appropriate parametric (de-)amplification factor. After being generated in the SHG, the pump field is guided to a ring-cavity. This mode cleaner has a Finesse of 555 and a linewidth (FWHM) of 1,3\,MHz. The purpose of this resonator is to attenuate high frequency phase noise, which is inherent on the green beam due to RF-phase-modulation used for locking the SHG cavity length. It has been shown that phase noise on the green pump beam can deteriorate the squeezing strength \cite{Furus07,FHDFS06}.
Utilizing the photo diode PD$_{MC532}$ an error signal for the cavity length control is generated via the Pound-Drever-Hall technique at a modulation frequency of 120\,MHz, which is much higher compared to the cavity linewidth.

\textit{Squeezing resonator length control beam} - A second NPRO-laser source (200\,mW, Innolight Mephisto OEM Product Line), which is frequency locked to the main laser on the squeezing breadboard, serves for the cavity length control of the squeezed light source. Using the orthogonal polarization (p-polarization) for locking the cavity length, no technical laser noise is introduced into the squeezed beam at the fundamental frequency ($\omega_0$, s-polarization).
Due to the birefringence of the nonlinear crystal inside the squeezed light source, however, this coherent control beam has to be frequency shifted to be simultaneously resonant with the generated s-polarized squeezed field. The frequency offset was determined to approximately 12.6\,MHz. This was measured while injecting both the alignment beam at frequency $\omega_0$ and the frequency shifted control beam. While scanning the cavity length, both orthogonally polarized TEM$_{00}$ Airy-peaks had to be overlapped. For monitoring the simultaneously resonant Airy-peaks, the photodetector PD$_{Alignment}$ was used.

\textit{Squeezing angle control beam} - A third NPRO-laser (again with an output power of 200\,mW) is set up for coherent control of the squeezing ellipse orientation with respect to the diagnostic homodyne detector. For this it is sufficient to inject only 25$\mu$W  into the squeezed light source cavity in order to generate clear error signals. This auxiliary laser is again frequency locked to the main laser using a PLL with an offset frequency of 15.2\,MHz. Operating on the GEO site, this coherent control field will finally be used to stabilize the phase relation between the squeezed vacuum field and the GEO\,600 interferometer signal output field. Furthermore, this coherent control field will be used to set up an auto alignment system for the squeezed beam into the interferometer. Two locking loops are set up for  full coherent control of the squeezed vacuum beam using this coherent control beam. The first loop stabilizes the relative phase between the injected control beam and the green pump field. To this end, an error signal is obtained by detecting a fraction of the (frequency shifted) light, which is back-reflected from the squeezed light source. Thus the photocurrent of the photodetector PD$_{GreenPhase}$ has to be demodulated at \emph{twice} the offset frequency \cite{VCHFDS06,VCDS07}. The feedback is applied to a phase shifter device in the green pump path with a unity gain frequency of 6\,kHz. A second control loop is set up in order to stabilize the phase relation between the squeezed vacuum field and the local oscillator beam of the homodyne detector. The error signal is extracted from the subtracted photocurrents of both homodyne photo diodes via RF-demodulation at the coherent control beam offset frequency. A phase shifter in the local oscillator beam path serves as an actuator. The locking bandwidth of this control loop is about 10\,kHz.

%%%%%%%%%%%%%%%%%%%%%%%%%%%%%%%%%%%%%%%%%%%%%%%%%%%%%%%%%%%%%%%%%%%%%%%%%%%%%%%%
%%%%%%%%%%%%%%%%%%%%%%%%%%%%%%%%%%%%%%%%%%%%%%%%%%%%%%%%%%%%%%%%%%%%%%%%%

\section{Results and discussion}

\begin{figure}[t]
\centerline{\includegraphics[width=0.8\textwidth,keepaspectratio,angle=0]{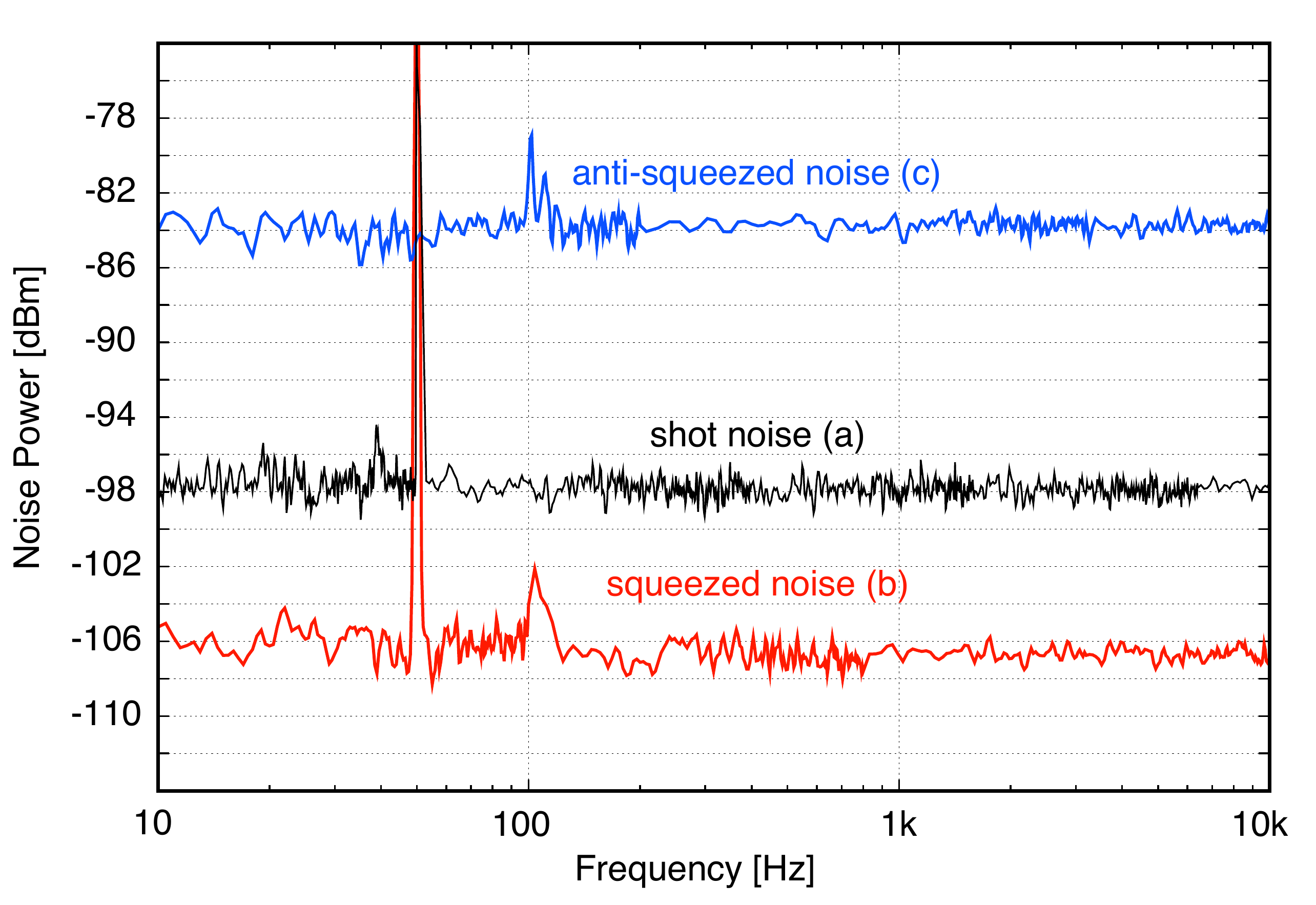}}
\caption{Quantum noise measurements performed with a balanced homodyne detector (BHD) using a 500\,$\mu$W local oscillator power. Trace (a) constitutes the shot-noise (vacuum noise) reference of the BHD, measured with the squeezed light input blocked. Trace (b) shows the observed squeezed quantum noise from our source. A nonclassical noise suppression of up to 9\,dB below shot-noise (a) was measured throughout the complete spectrum from 10\,Hz up to 10\,kHz. The corresponding anti-squeezing (c) was 14\,dB above the shot-noise level. The electronic dark noise (not shown) was 17\,dB below the shot-noise and was not subtracted from the measured data. The peaks at 50\,Hz and 100\,Hz were due to the electric mains supply.}
\label{SqueezingMeasurement}
\end{figure}

In Figure \ref{SqueezingMeasurement}, quantum noise measurements of the diagnostic balanced homodyne detector in the frequency band from 10\,Hz up to 10\,kHz are shown. Trace (a) represents a shot-noise measurement performed with the signal input port blocked of the homodyne detector. The local oscillator intensity was 500\,$\mu$W throughout all the measurements. This LO power delivered a large dark noise clearance of 17\,dB. 
%The electronic dark noise is not shown in Figure \ref{SqueezingMeasurement}.
When the squeezed states from our source are mode-matched into the homodyne detector signal input port the noise level changes according to the quantum noise variance of these states. Trace (b) shows the variance of the squeezed quantum noise and trace (c) the variance of the anti-squeezed noise. In order to perform these measurements, the local oscillator phase was stabilized to either the squeezed or the anti-squeezed field quadrature of the signal input, respectively. For the upgrade of GEO\,600 one is obviously interested in the stabilization of the phase between the squeezed field and the GEO\,600 signal field in such a way that the finally detected quantum noise is squeezed. Nevertheless, a full characterization of the squeezed field is only possible through the measurement of both the squeezed and anti-squeezed quadrature field variances. The squeezing level achieved in the detection band from 10\,Hz to 10\,kHz is at least 8\,dB below the shot-noise level. At several frequencies, squeezing of up to 9\,dB is observed. The corresponding anti-squeezing level was measured to be 14\,dB above the shot-noise level over the entire bandwidth (trace (c)). From these measurements a total optical loss of approximately 10\,\% on the squeezed laser field can be derived. This loss value \textit{includes} the homodyne detection efficiency of approximately 95\,\%.
Thus, if the squeezed states are directly guided into the signal output port of GEO\,600, the diagnostic homodyne detector loss can be subtracted, and more than 10\,dB squeezing will be injected into the GW-detector.

In the following we estimate the expected nonclassical sensitivity improvement of the future squeezed-light enhanced GEO\,600 detector assuming that additional optical loss will reduce the squeezing strength. We consider frequencies at which GEO\,600 is currently shot-noise limited, i.e. above a few hundred Hertz. 
We estimate the additional optical loss for the squeezed field to 10\,\% up to 15\,\%. Here, we consider (i) a non-perfect mode matching to the GEO\,600 signal-recycling cavity, (ii) loss in the input Faraday isolator, (iii) non-perfect dielectric coatings of GEO\,600 optics, and (iv) the non-perfect photo-diode quantum efficiency.  From these assumptions we conclude, that finally a 6\,dB nonclassical sensitivity improvement of GEO\,600 might be reachable. This number corresponds to a sensitivity improvement that is equivalent to an increase in laser power by a factor of 4, without however, the unwanted side-effects from a higher thermal load.

\section{Conclusion}

We presented a detailed description of the optical setup of the GEO\,600 squeezed light source and showed first measurement results. Up to 9\,dB of squeezing over the entire bandwidth of the earth-based gravitational wave detectors was demonstrated. To the best of our knowledge, this is the highest measured squeezing value at audio frequencies observed so far. Our result also belongs to the highest squeezing values ever measured. At radio frequencies (MHz) only recently slightly higher values between 9\,dB and 11.5 dB were reported \cite{VMCHFLGDS08,Furus07,Moritz11dB}. We have estimated the additional optical loss for the squeezed-light field when injected into GEO\,600 and come to the conclusion that a non-classical detector sensitivity improvement of 6\,dB might be possible for the shot-noise limited band of GEO\,600. After a long-term test the squeezed light source will be ready for the implementation in GEO\,600.

\ack{This work has been supported by the Deutsche Forschungsgemeinschaft (DFG) through Sonderforschungsbereich 407 and the Centre for Quantum Engineering and Space-Time Research QUEST.}

%%%%%%%%%%%%%%%%%%%%%%%%%%%%%%%%%%%%%%%%%%%%%%%%%%%%%

\end{document}